\documentclass{elsart}

 \usepackage{graphicx}

\usepackage{amssymb}

\begin{document}

\begin{frontmatter}



\title{Monte Carlo simulations of a tethered membrane model on a disk with intrinsic curvature}

\author[label1]{Motonari Igawa}, 
\author[label2]{Hiroshi Koibuchi}, and
\ead{koibuchi@mech.ibaraki-ct.ac.jp}
\author[label1]{Mitsuru Yamada}

\address[label1]{Department of Mathematical Sciences, Faculty of Sciences, Ibaraki University, 
Bunkyo 2-1-1, Mito, Ibaraki 310-8512}
\address[label2]{Department of Mechanical and Systems Engineering, Ibaraki College of Technology, 
Nakane 866, Hitachinaka,  Ibaraki 312-8508, Japan}

\begin{abstract}
A first-order phase transition separating the smooth phase from the crumpled one is found in a fixed connectivity surface model defined on a disk. The Hamiltonian contains the Gaussian term and an intrinsic curvature term. 
\end{abstract}

\end{frontmatter}

\section{Introduction}\label{intro}
There has been a renewal of interest in membrane physics \cite{NELSON-SMMS-2004} of Helfrich and Polyakov-Kleinert \cite{HELFRICH-NF-1973,POLYAKOV-NPB-1986,Kleinert-PLB-1986}. The phase transition between the smooth phase and the crumpled phase is one of the interesting topics in the surface model \cite{DavidGuitter-EPL-1988,Peliti-Leibler-PRL-1985,BKS-PLA-2000,BK-PRB-2001,Kleinert-EPJB-1999,BOWICK-TRAVESSET-PREP-2001,WIESE-PTCP19-2000,WHEATER-JP-1994}. The smooth (crumpled) phase is appeared in the model with infinite (zero) bending rigidity $b$ which is included in the Hamiltonian $S$ such that $S\!=\!S_1\!+\!bS_2$, where $S_1$ and $S_2$ are the Gaussian energy and the bending energy respectively.

We will concentrate on the phantom and tethered surface model in this Letter. Models on triangulated surfaces can be classified into two groups. One of them is the fluid model which is defined on dynamical connectivity surfaces \cite{CATTERALL-NPB-SUPL-1991,AMBJORN-NPB-1993,BCHGM-NPB-1993,ABGFHHE-PLB-1993,KOIB-PLA-2002,KOIB-PLA-2004,KOIB-PLA-2003-2,KOIB-EPJB-2004}, and the other is the tethered model on fixed connectivity surfaces \cite{KOIB-EPJB-2004,BCFTA-1996-1997,WHEATER-NPB-1996,KANTOR-NELSON-PRA-1987,GOMPPER-KROLL-PRE-1995,KOIB-PRE-2004-1,KOIB-PRE-2004-2,KANTOR-SMMS-2004,KANTOR-KARDER-NELSON-PRA-1987,BCTT-EPJE-2001,BOWICK-SMMS-2004}. There is another classification of surface models; a model is called real or phantom according to whether the surface is self-avoiding or not. The phantom model is still interesting because the self-avoiding interaction is considered to be irrelevant in the limit $N\!\to\!\infty$ \cite{NELSON-SMMS-2004-2}.  

We consider that the phase structure of the tethered surface model remains to be studied further. It has been recognized that the phantom tethered model undergoes a second-order phase transition both on a sphere \cite{KOIB-EPJB-2004,BCFTA-1996-1997,WHEATER-NPB-1996,KANTOR-NELSON-PRA-1987,GOMPPER-KROLL-PRE-1995,KOIB-PRE-2004-1,KOIB-PRE-2004-2,KANTOR-SMMS-2004,KANTOR-KARDER-NELSON-PRA-1987} and on a disk \cite{BCTT-EPJE-2001,BOWICK-SMMS-2004}. On the other hand, it has also been reported that there is a first-order transition in the model with Hamiltonian slightly different from the ordinary one of Helfrich and Polyakov-Kleinert on a sphere \cite{KOIB-PRE-2004-1}. First-order transitions can also be seen in a model of Nambu-Goto Hamiltonian with a deficit angle term \cite{KOIB-PRE-2004-2} and in a model with Hamiltonian containing the Gaussian term and an intrinsic curvature term \cite{KOIB-EPJB-2004}. Therefore, phase transitions of the tethered surface model should be understood more clearly.
 
In this article, we would like to investigate the dependence of the phase transition on the existence of boundary in the surface. For this purpose, we define the model on a triangulated disk, which is topologically identical with a sphere with a hole, The disk is obtained by dividing the hexagon. The Hamiltonian is given by a linear combination of the Gaussian term and an intrinsic curvature term; intrinsic curvatures in membranes have no direct analog in one dimensional objects like linear polymer as noted in \cite{NELSON-SMMS-2004}. It was reported that the model with intrinsic curvature undergoes a first-order transition on a sphere \cite{KOIB-EPJB-2004} as mentioned above. Hence, it is interesting to see whether the boundary influences the first-order phase transition of the model.  

\section{The model}\label{model}
\begin{figure}[hbt]
\centering
\includegraphics[width=6cm,height=5.4cm,]{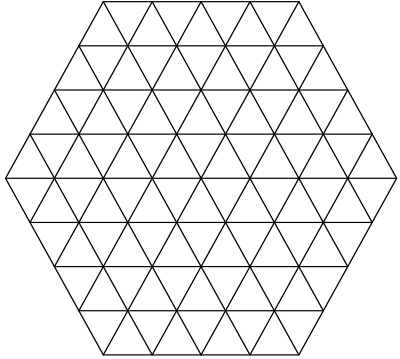}
\caption{A triangulated disk of size $(N,N_E,N_T)\!=\!(61,156,96)$, where $N,N_E,N_T$ are the total number of vertices, the total number of edges, the total number of triangles.}

\label{fig-1}
\end{figure}
The model is defined on a triangulated disk which is obtained by dividing the hexagon. Figure \ref{fig-1} is a lattice of size $N\!=\!61$, which is obtained by dividing every edge of the original hexagon into $4$-pieces. If the edges are divided by $L$-pieces we have a lattice of $(N,N_E,N_T)\!=\!(3L^2\!+\!3L\!+\!1, 9L^2\!+\!3L,6L^2)$, where $N,N_E,N_T$ are the total number of vertices, the total number of edges, and the total number of triangles.

The Gaussian tethering potential $S_1$ and the intrinsic curvature $S_3$ are defined by
\begin{equation}
\label{S3}
S_1=\sum_{(ij)} \left(X_i-X_j\right)^2, \; S_3= \sum_i \left( \delta_i - \delta_0    \right)^2, 
\end{equation}
where $\sum_{(ij)}$ is the sum over bond $(ij)$ connecting the vertices $X_i$ and $X_j$, and  $\delta_i$ in $S_3$ is the sum of angles of the triangles meeting at the vertex $i$. The value of the symbol $\delta_0$ in $S_3$ is given by 
\begin{equation}
\label{delta-0}
\delta_0=\left\{ \begin{array}{ll}
    2\pi,                   & \mbox{(internal vertices);} \\
    \pi (\mbox{or}\; 2\pi/3), & \mbox{(boundary vertices).} 
   \end{array} \right.
\end{equation}
$2\pi/3$ is assigned to the six vertices, and $\pi$ to the remaining vertices on the boundary. The definition of $\delta_0$ in (\ref{delta-0}) allows us to call $S_3$ as a deficit angle term. 

The partition function is defined by 
\begin{eqnarray}
 \label{partition-function}
Z(\alpha) = \int \prod _{i=1}^N dX_i \exp\left[-S(X)\right],\qquad \\
S(X)=S_1 + \alpha S_3,  \qquad\quad \nonumber
\end{eqnarray}
where $N$ is the total number of vertices as described above. The expression $S(X)$  shows that $S$ explicitly depends on the variable $X$. The coefficient $\alpha$ is a modulus of the intrinsic curvature. The surfaces are allowed to self-intersect. The center of surface is fixed in $Z(\alpha) $ to remove the translational zero mode.

 It should also be noted that the Gaussian tethering potential $S_1$ at the internal vertices differs from the one at the boundary vertices because the co-ordination number of them is different from each other. The intrinsic curvature $S_3$ at the internal vertices can also be different from those at the boundary one on fluctuating surfaces due to the same reason for $S_1$.  These differences of $S_1$ and $S_3$ between the internal and the boundary vertices are the one we can not find in compact surfaces such as a sphere.

We comment on a relation of $S_3$ in (\ref{S3}) to an intrinsic curvature energy $S_3\!=\!-\sum_i \log(\delta_i/2\pi)$ in \cite{KOIB-EPJB-2004}. This $S_3$ in  \cite {KOIB-EPJB-2004} is intimately related with the co-ordination dependent term  $S_3\!=\!-\sum_i \log (q_i/6)$, which comes from the integration measure $\prod_i dX_i q_i^\alpha$ \cite{DAVID-NPB-1985} in the partition function for the model on a sphere. $S_3\!=\!-\sum_i \log(\delta_i/2\pi)$ becomes minimum if $\delta_i\!=\!2\pi$ for all $i$, and hence it becomes smaller on a smooth sphere than on a crumpled one. However, the flat configuration of the hexagonal lattice shown in Fig. \ref{fig-1} does not minimize this $S_3$, because there are vertices such that $\delta_i\!\not=\!2\pi$. This is the reason why we use $S_3$ in (\ref{S3}), which becomes zero on such flat configuration as shown in Fig. \ref{fig-1}. 

\section{Monte Carlo technique}\label{MC-Techniques}
The canonical Monte Carlo technique is used to update the variables $X$ so that $X^\prime \!=\! X \!+\! \delta X$, where the small change $\delta X$ is made at random in a small sphere in ${\bf R}^3$. The radius $\delta r$ of the small sphere is chosen at the beginning of the simulations to maintain the rate of acceptance $r_X$ for the $X$-update as $0.4 \leq r_X \leq 0.6$. 

We use a random number called Mersenne Twister \cite{Matsumoto-Nishimura-1998} in the MC simulations. Two sequences of random number are used; one for 3-dimensional move of vertices $X$ and the other for the Metropolis accept/reject in the update of $X$.

 We use surfaces of size $N\!=\!2107$, $N\!=\!3997$, $N\!=\!6487$, and  $N\!=\!9577$, which correspond to  the partitions $L\!=\!26$, $L\!=\!36$, and  $L\!=\!56$ respectively.  The size of the surfaces is relatively larger than that of spherical surfaces used in \cite{KOIB-EPJB-2004}.

The total number of MCS after the thermalization MCS is about $1.8\!\times\!10^8$ at the transition point of surfaces of size $N\!=\!9577$,  $N\!=\!6487$, and   $N\!=\!3997$,  and $1.6\!\times\!10^8$ for  $N\!=\!2107$. Relatively smaller number of MCS ($1.0\!\times\!10^8\sim 1.6\!\times\!10^8$) is done at $\alpha$  distant from the transition point.

\section{Results}\label{Results}
\begin{figure}[hbt]
\centering
\includegraphics[width=12cm]{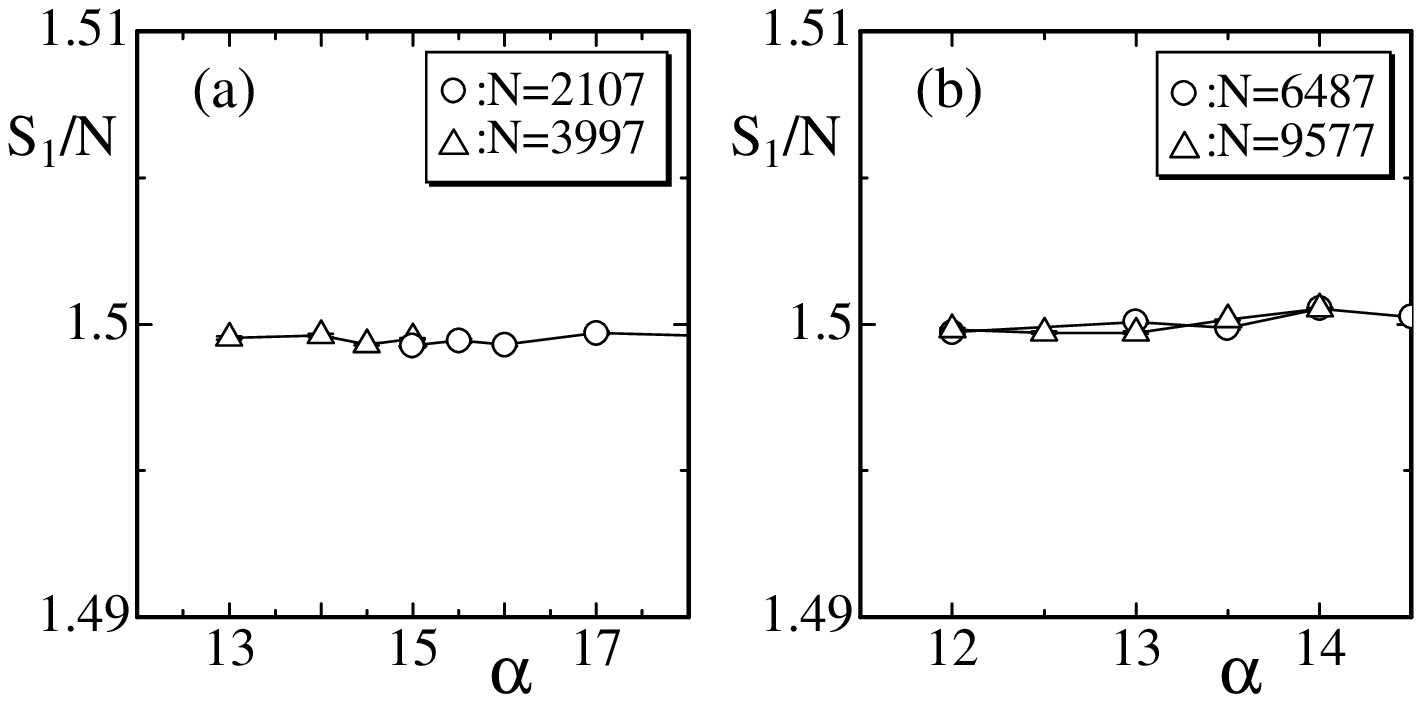}
\caption{$S_1/N$ vs. $\alpha$ obtained on surfaces of (a) $N\!=\!2107$, $N\!=\!3997$, and (b) $N\!=\!6487$, $N\!=\!9577$. }
\label{fig-2}
\end{figure}

Scale invariance of the partition function predicts that $S_1/N\!=\!1.5$, which is not influenced by the boundary condition imposed on the model. We also expect that this relation should not be influenced by whether the phase transition is of first order or not.  Figures \ref{fig-2} (a) and \ref{fig-2} (b) are $S_1/N$ against $\alpha$ obtained on surfaces of size $N\!=\!2107$, $N\!=\!3997$ and $N\!=\!6487$, $N\!=\!9577$. We find from these figures that the expected relation $S_1/N\!=\!1.5$ is satisfied. 

\begin{figure}[hbt]
\centering
\includegraphics[width=12cm]{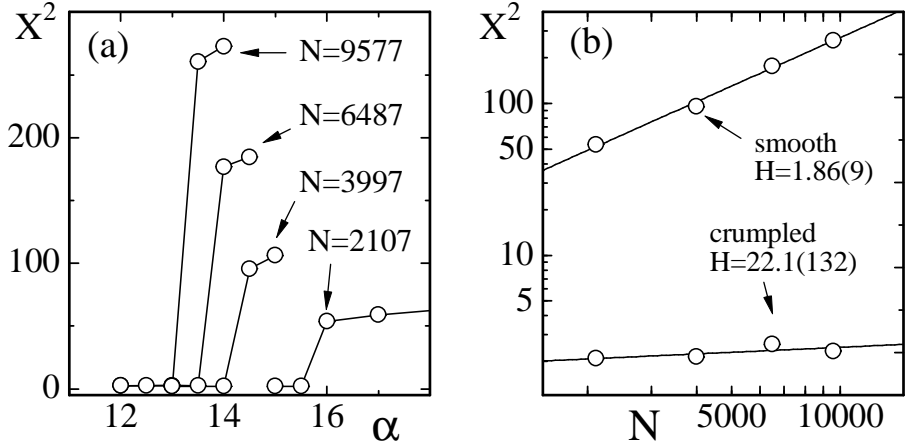}
\caption{(a) $X^2$ vs. $\alpha$ obtained on surfaces of size from $N\!=\!2107$ to $N\!=\!9577$, and (b) log-log plots of $X^2$ vs. $N$ obtained just below and above the transition point $\alpha_c$ on each surface. The straight lines in (b) are drawn by fitting the data to Eq. (\ref{Hausdorff}).}
\label{fig-3}
\end{figure}

We expect that the surfaces become flat at $\alpha\!\to\!\infty$ and crumpled at $\alpha\!\to\!0$. Hence the size of the surfaces can be reflected in the mean square size $X^2$ defined by
\begin{equation}
\label{X2}
X^2={1\over N} \sum_i \left(X_i-\bar X\right)^2, \quad \bar X={1\over N} \sum_i X_i.
\end{equation}
In order to see the size of surfaces, we plot $X^2$ against $\alpha$ in Fig. \ref{fig-3} (a). We see that $X^2$ discontinuously changes at finite $\alpha$. This indicates a discontinuous transition between the smooth phase and the crumpled one. 

Figure \ref{fig-3} (b) shows log-log plots of $X^2$ against $N$, obtained just below and above the transition point $\alpha_c$ where $X^2$ discontinuously changes as shown in Fig. \ref{fig-3} (a). The straight line denoted by smooth (crumpled) is obtained by fitting $X^2$ in the smooth (crumpled) phase just above (below) $\alpha_c$ by
\begin{equation}
\label{Hausdorff}
X^2 \sim N^{2/H},
\end{equation}
where $H$ is the Hausdorff dimension. Thus we have 
\begin{equation}
\label{H}
H=\left\{ \begin{array}{ll}
    1.86\pm 0.09,                   & (\mbox{smooth}); \\
   22.1\pm 13.2, & (\mbox{crumpled}). 
   \end{array} \right.
\end{equation}
The result $H\!=\!1.86(9)$ in (\ref{H}) implies that the surface is almost smooth in the smooth phase. On the contrary,  $H\!=\!22.1(132)$ implies that the surface is highly crumpled in the crumpled phase. 

\begin{figure}[hbt]
\centering
\includegraphics[width=12cm]{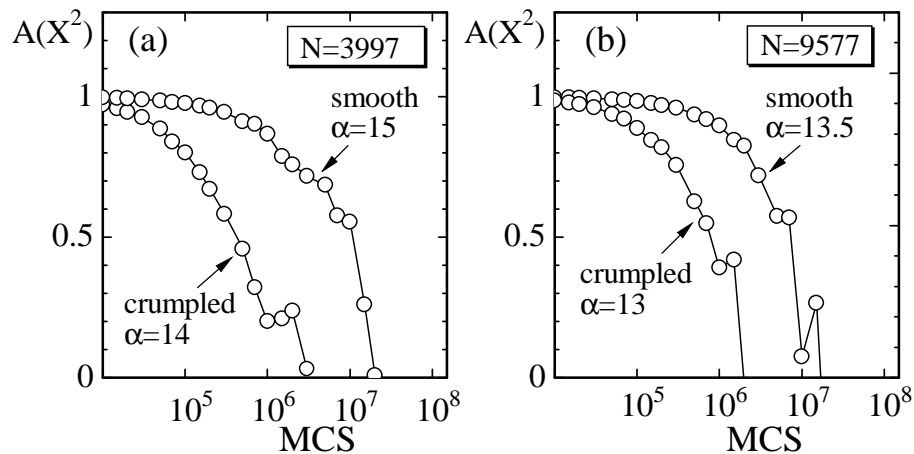}
\caption{Auto-correlation coefficient $A(X^2)$ in the smooth and the crumpled phases on surfaces of size (a) $N\!=\!3997$, and (a) $N\!=\!9577$.}
\label{fig-4}
\end{figure}
The phase transition can be reflected in the convergence speed of the MC simulations. In order to see this, we plot in Figs. \ref{fig-4}(a) and \ref{fig-4}(b) the autocorrelation coefficient $A(X^2)$ of $X^2$ defined by 
\begin{eqnarray}
A(X^2)= \frac{\sum_i X^2(\tau_{i}) X^2(\tau_{i+1})} 
   {  \left[\sum_i  X^2(\tau_i)\right]^2 },\\ \nonumber
 \tau_{i+1} = \tau_i + n \times 500, \quad n=1,2,\cdots.  
\end{eqnarray}
The horizontal axes in the figure represent $500\!\times\! n\;(n\!=\!1,2,\cdots)$-MCSs, which is a sampling-sweeps between the samples $X^2(\tau_i)$ and $X^2(\tau_{i+1})$ . The coefficients $A(X^2)$ in Figs. \ref{fig-4}(a) and \ref{fig-4}(b) are obtained on surfaces of size $N\!=\!3997$ and $N\!=\!9577$ respectively. We clearly see from the figures that the convergence speed in the smooth phase is about 10 times larger than that in the crumpled phase. The reason why the convergence speed in the smooth phase is so larger than in the crumpled phase is that the phase space volume ($\subseteq {\bf R}^3$), where the vertices $X_i$ take their values, is relatively larger in the smooth phase than that in the crumpled phase. This difference in the phase space volume can also be reflected in $X^2$ shown in Figs. \ref{fig-3}(a) and \ref{fig-4}(b). 

However, we find no phenomena of the critical slowing down in $A(X^2)$ at the transition point. The reason of this is because of irreversibility of the transition. The change from the crumpled phase to the smooth one appears to be irreversible. If the vertices once localize to a small region in ${\bf R}^3$ at the transition point, they hardly expand to be a smooth surface. This irreversibility was also seen in the same model on a sphere \cite{KOIB-EPJB-2004}. 

\begin{figure}[hbt]
\centering
\includegraphics[width=12cm]{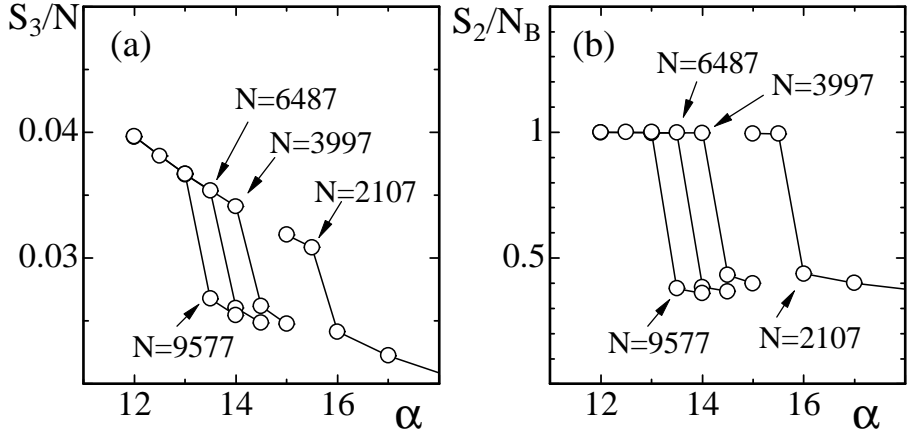}
\caption{(a) $S_3/N$ vs. $\alpha$, and (b) $S_2/N_B$ vs. $\alpha$, where $N_B(=\!N_E\!-6L)$ is the total number of internal bonds.}
\label{fig-5}
\end{figure}
Figure \ref{fig-5}(a) shows the intrinsic curvature energy $S_3/N$ changing discontinuously against $\alpha$. This indicates the existence of a first-order phase transition. The variation of $S_2/N_B$ against $\alpha$ is shown in Fig. \ref{fig-5}(b), where $S_2$ is the bending energy defined by $S_2\!=\!\sum_i(1-\cos\theta_i)$ and $N_B$ is the total number of internal bonds. It should be noted that $6L$-bonds (the total number of bonds on the boundary) are not included in $N_B(=\!N_E\!-6L)$. The quantity $1-\cos\theta_i$ is defined only on the internal bonds. The discontinuous change of $S_2/N_B$ shown in Fig. \ref{fig-5} (b) indicates that the phase transition separates the smooth phase from the crumpled phase. 

\begin{figure}[hbt]
\centering
\includegraphics[width=12cm]{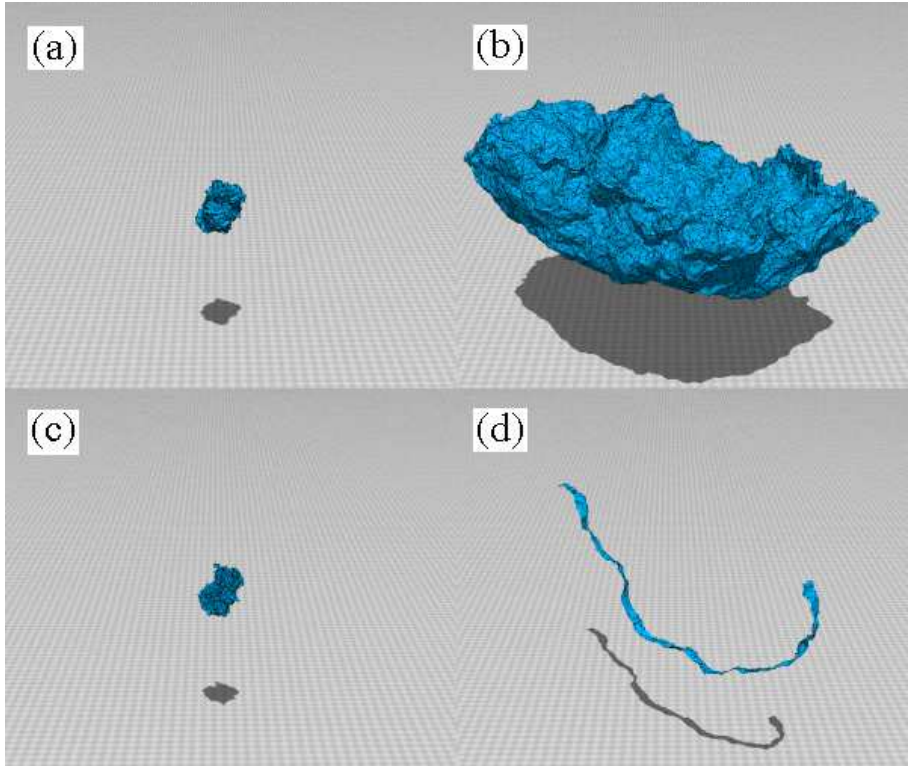}
\caption{Snapshots of surfaces of size $N\!=\!9577$ at (a) $\alpha\!=\!13$, and at (b) $\alpha\!=\!13.5$, and (c) a section of the surface in (a), (d) a section of the surface in (b).  Figures are shown in the same scale. }
\label{fig-6}
\end{figure}
Snapshots of surfaces are shown in Figs. \ref{fig-6}(a) and \ref{fig-6}(b), which are obtained at $\alpha\!=\!13$ and $\alpha\!=\!13.5$. The size is $N\!=\!9577$ in Figs. \ref{fig-6}(a) and \ref{fig-6}(b). Sections of the surfaces in Figs. \ref{fig-6}(a) and \ref{fig-6}(b) are shown in Figs. \ref{fig-6}(c) and \ref{fig-6}(d) respectively. It is easy to see that the surface in Fig. \ref{fig-6}(a) is crumpled and that the surface in Fig. \ref{fig-6}(b) is smooth. Although the surface in Fig. \ref{fig-6}(b) looks like a cup, it is almost smooth as can be seen from the section in Fig. \ref{fig-6}(d). 

\section{Summary and conclusion}\label{Conclusions}
We have shown that a tethered surface model with an intrinsic curvature undergoes a first-order phase transition between the smooth phase and the crumpled phase on a disk. Our results indicate that there is no influence of the boundary condition on the phase transition of the model with intrinsic curvature. 

The Gaussian tethering potential and the intrinsic curvature energy are defined on the triangulated disk which is obtained by dividing the edges of the hexagon into $L$-pieces. The triangulated disk is characterized by $(N,N_E,N_T)\!=\!(3L^2\!+\!3L\!+\!1, 9L^2\!+\!3L,6L^2)$, where $N_E,N_T$ are the total number of edges and the total number of triangles. We used surfaces of size up to $N\!=\!9577$, which is about four times larger than those used in \cite{KOIB-EPJB-2004} for the similar model on a sphere. Hence, the observed first-order transition in this Letter is considered as the one in the thermodynamic limit of the model. The canonical Monte Carlo technique was used to update the position variable $X$ of vertices. 

Combining the result in this Letter and those in \cite{KOIB-EPJB-2004}, we can conclude that 
\begin{enumerate}
\item [1)] the tethered surface model with intrinsic curvature undergoes a first-order phase transition
\item [2)] the phase transition can be observed both on a sphere and on a disk. 
\end{enumerate}

It is also interesting to study the phase structure of the tethered model on a torus, and that of the fluid model.

This work is supported in part by a Grant-in-Aid for Scientific Research, No. 15560160.



\end{document}